# Non-reciprocal coherent all-optical switching between magnetic multi-states

T. Zalewski[1], V. Ozerov[1], A. Maziewski[1], I. Razdolski[1] & A. Stupakiewicz[1]*

[1]Faculty of Physics, University of Bialystok, 1L Ciolkowskiego, 15-245 Bialystok, Poland.

*Correspondence to: and@uwb.edu.pl (A.S.).

**Abstract.** We present experimental and computational findings of the laser-induced non-reciprocal motion of magnetization during ultrafast photo-magnetic switching in garnets. We found distinct coherent magnetization precession trajectories and switching times between four magnetization states, depending on both directions of the light linear polarization and initial magnetic state. As a fingerprint of the topological symmetry, the choice of the switching trajectory is governed by an interplay of the photo-magnetic torque and magnetic anisotropy. Our results open a plethora of possibilities for designing energy-efficient magnetization switching routes at arbitrary energy landscapes.

**Introduction**

Understanding the interplay between topology, crystal symmetry, and free energy landscape map is key for the manipulation of magnetic moments in complex systems. Control of spin dynamics at ultrashort timescales by various stimuli has recently become a vibrant research direction revolving around physical mechanisms acting on magnetization as well as their dynamical characteristics[1–4]. Steering the magnetization along energy-efficient routes across the magnetic anisotropy landscape requires the highly sought-after ability to control the torque. Multiple mechanisms have been suggested, including current modulation in spin-orbit torque systems[5,6], employing ultrashort acoustic pulses[7–9], and utilizing coherent phonon-magnon coupling[10–13]. In general, among the variety of methods, the ability to manipulate magnetization solely using laser pulses holds tremendous potential for future technologies, facilitating the fastest-ever data recording with minimal heat dissipation. On top of essentially thermal mechanisms of all-optical magnetization switching[14,15], of particular interest is the non-thermal photo-magnetic excitation where spin-orbit interaction mediates the modification of the spin energy landscape through absorption of optical radiation. Over the last decades, a solid body of knowledge has been accumulated aiming at bringing the photo-magnetism onto ultrafast timescale and separating it from concomitant thermal and inverse magneto-optical effects[16–21]. Owing to its non-thermal nature, the dynamics of photo-magnetism, in which magnetization is not quenched upon pulse laser irradiation, heavily relies on the magnetization precession. The latter is usually described by the Landau-Lifshitz-Gilbert (LLG) dynamics with a time-dependent effective magnetic field[22]. This highlights the importance of the topography of a magnetic landscape, which could play a key role in determining the magnetization switching conditions as well as the relevant dynamical characteristics. Materials with a cubic magnetic symmetry usually have a large number of degenerate energy minima, and smaller angles between them facilitate the nonlinear regime of large-angle magnetization precession[23] and eventually all-optical switching[24].

Since the magnetization is set into motion through an ultrafast modification of the energy landscape, it is of utter importance to understand the optical response of the latter in detail. This dynamical topography can then be used advantageously, for example, reducing the energy consumption or accelerating the switching dynamics. The latter is particularly relevant in light of the recently demonstrated controllable magnetization reversal between two equilibrium states at the picosecond timescale[24]. The open questions that we address in this work pertain to the detailed analysis of the trajectories of the magnetization switching as well as the anatomy of the photo-magnetic excitation in high-symmetry media. Employing time-resolved magneto-optical imaging technique, we get valuable insights into the coherent dynamics of simultaneous switching of coexisting magnetic states by a single laser pulse. We demonstrate the existence of two non-equivalent switching trajectories in cubic photo-magnetic yttrium-iron garnet with Co-ions (YIG:Co), whereas the forward and backward routes of magnetization can be chosen with the polarization of light. We further highlight the key role of an interplay between the torque and magnetic symmetries in the laser-induced motion of spins in co-existing magnetic domains. We argue that the apparent (almost twofold) difference in the switching times between the two pairs of states represents another fingerprint of the torque symmetry exhibiting coherent magnetization switching.

**Results**

**Time-resolved multi-states magnetization switching.** The photo-magnetic YIG:Co(001) film is characterized by the predominantly cubic magnetic anisotropy with eight energy minima, with magnetization oriented close to the diagonals of the cubic unit cell[25,26] (see Supplemental Material). Among those minima, we focus on four (Fig. 1a) which exhibit the full richness of the magnetization dynamics upon photo-magnetic excitation with a single 50 fs pump laser pulse. To achieve the highest photo-magnetic efficiency of the switching[27], the pump wavelength was set to 1300 nm (see SM). In contrast to our previous observations[24], here we employed the highly-sensitive time-resolved single-shot magneto-optical imaging technique[28] which allows for simultaneous observation of magnetization dynamics in coexisting magnetic phases and retrieve both the spatial and temporal dynamics of multi-domain states. In particular, we monitored the laser-induced dynamics of the magnetization in domains depicted in Fig. 1a through transient variations of the Faraday rotation of the defocused probe beam at 650 nm wavelength.

To improve the signal-to noise ratio, background images with the pump beam blocked were subtracted at each time delay, thus creating the presented stack of differential images of the normal [001] magnetization projection $M_z$. Selecting regions of interest within the borders of individual domains and integrating the variations over them, we obtained a set of traces corresponding to the laser-induced $\Delta M_z$ dynamics in each of the magnetic states as a function of the optical delay between the pump and probe pulses $\Delta t$ (see SM).

In the image sequence complemented by (or together with) a sketch of magnetization states in Fig. 1b and 1c, we show that the magnetization switching occurs in both large ($M_1$ and $M_8$) and small ($M_4$ and $M_5$) domains simultaneously. We note that the pattern of the small domains remains constant during the entire experiment, thus confirming the non-thermal nature of the photo-magnetic switching. Interestingly, after about 30 ps the magnetization starting its dynamics from the $M_1$ direction obtains the transient perpendicular state to the sample plane. This is in contrast with the previously reported[24] magnetization switching between the canted states in YIG:Co via the in-plane direction.

A typical example of the experimental $\Delta M_z(\Delta t)$ traces in the two states with opposite normal magnetization components is shown in Fig. 1d, illustrating the switching along the $M_8 \rightarrow M_1$ and $M_4 \rightarrow M_5$ routes with the $E \parallel [100]$ pump pulses. It is seen that the trajectories are clearly asymmetric. The sketch in Fig.1c illustrates the rotation of the magnetization pair in adjacent domains (red and blue arrows), where the transient variations of the angle between them are inextricably linked to the inequality of the two trajectories.

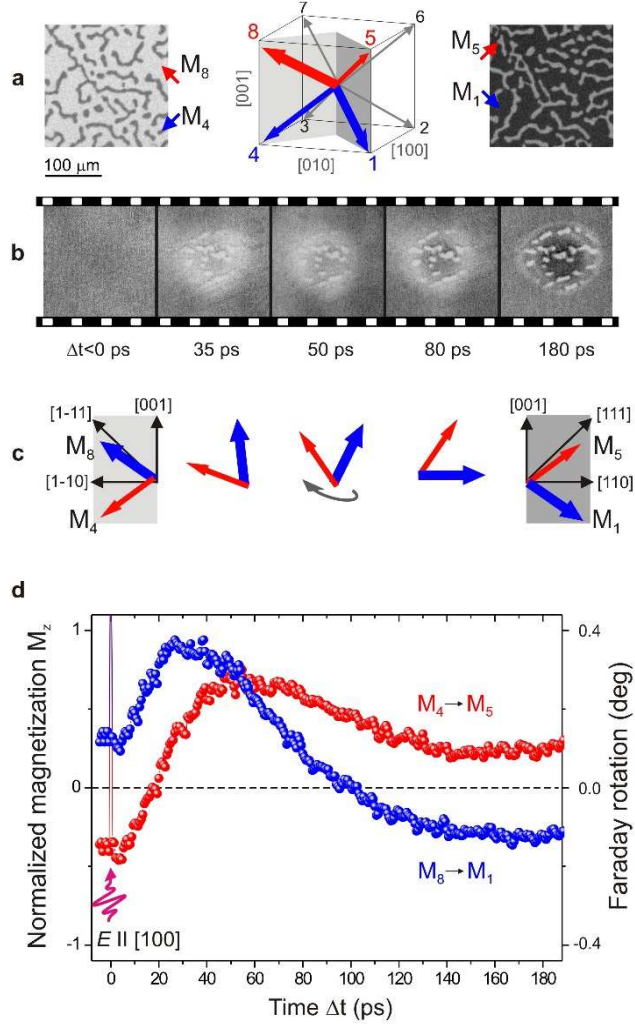

**Figure 1. Multi-states magnetization switching in a cubic magnetic system. a** Images of multi-states magnetic domain structure and orientation of easy magnetization axes in YIG:Co films. **b** The differential stack of time-resolved images illustrating the transient dynamics of photo-magnetic switching in YIG:Co. The experiment involved the use of a 50 fs single pump pulse with a fluence of 50 mJ cm$^{-2}$ and polarization along $E \parallel [100]$ direction in the garnet. **c** the sketch illustrates the motion and rotation of magnetization at domains for different delay time $\Delta t$. **d** The normalized out-of-plane component of magnetization $M_z$ for the simultaneous switching between four magnetic states ($M_8$-$M_1$ and $M_4$-$M_5$).

Similar measurements performed when all 8 different magnetic states were prepared and illuminated with light at one of the two orthogonal polarizations ($E \parallel [100]$ and $E \parallel [010]$ directions) revealed that only two unique time traces of $\Delta M_z(\Delta t)$ can be observed (see SM). Thus, all switching trajectories can be assigned to either type I or type II, depending on the initial magnetization state and pump polarization. In Fig. 2a we show the averaged traces of both types for the orthogonal pump polarizations along the [100] and [010] directions, and the inequivalence of these trajectories becomes glaringly apparent. This observation is highly surprising in light of the fourfold crystalline symmetry and magnetic anisotropy of the garnet. To understand the physical origins of the dissimilarity of the two trajectories, we decomposed them into even and odd contributions. These are shown in Fig. 2b with green and black full data points, respectively. Notably, the even part closely resembles the averaged over multiple domains, a step-like switching trace reported previously[24]. It is seen that by 100 ps, the switching is already completed, and the remaining minute variations are related to the relaxation of magnetization within the corresponding potential minimum. On the other hand, the odd part exhibits a strongly damped oscillatory character, highlighting the precessional character of the switching dynamics in general. It is the odd contribution that is responsible for the non-reciprocal switching seen in Fig. 2a: the switching along the M₄→M₅ route and back proceeds along the different trajectories.

We argue that the odd, non-reciprocal contribution can be understood as a fingerprint of the interplay between the torque symmetry and the coherent magnetization dynamics. The solid line in Fig. 2b shows the (scaled) time derivative of the even contribution. Its striking reminiscence of the odd part suggests that the latter originates in the precession of magnetization when moving along the switching route. This results in the emergence of the orthogonal component $dM/dt$ since $(M \cdot dM/dt) = 0$. The sign of this orthogonal component is governed by the direction of movement along the switching route, thus giving rise to the odd contribution to the magnetization dynamics. In other words, a new curvilinear reference frame can be introduced where the even contribution depicting the genuine switching route is bereft of any rotational part. In such a reference frame, the LLG equation will be transformed accordingly, acquiring an additional term similarly to the mechanical equations of motion in a rotating reference frame. This additional term can be viewed as a fictitious magnetic field giving rise to the magnetization dynamics in the direction orthogonal to the switching route, or the aforementioned odd contribution. The characteristic shape of the normal projection of this contribution is then determined by the change rate of $M_z$ along the switching route, in agreement with Fig. 2b.

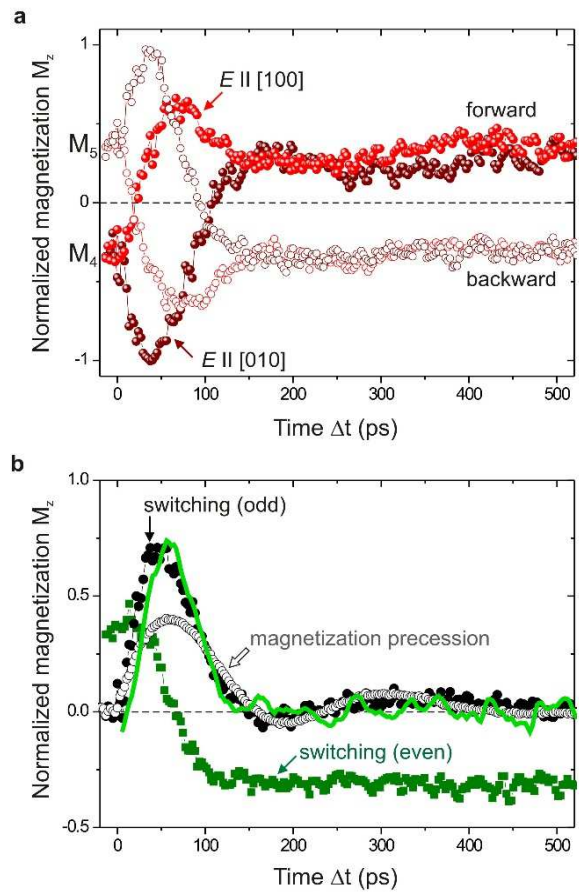

**Figure 2. Non-reciprocal of all-optical magnetization switching. a** Pump polarization effect on magnetization trajectory enabling selecting switching trajectory and final state for orthogonal orientation of laser pump polarization along [100] and [010] direction in YIG:Co film. **b** The differential signal of switched states (close black points) compared to the measured magnetization precession (open points) below the threshold level of switching. The even and odd contribution of photo-magnetic switching. The green solid line is the scaled derivative of the even component of photo-magnetic switching.

**Photo-magnetic torque symmetry.** To get further insight into the laser-induced asymmetry of the magnetization switching trajectories, we performed a tensor analysis of the photo-magnetic excitation along the lines discussed in[24]. In particular, the photo-magnetic Hamiltonian contribution reads:

$$\mathcal{H}_{p-m} = \hat{\beta} \mathbf{E} \mathbf{E}^* \mathbf{M} \mathbf{M}, \tag{1}$$

where $E$ is the electric field of the optical pump, and $\hat{\beta}$ is a fourth-order polar tensor responsible for the photo-magnetic susceptibility[27]. In the reference frame aligned with the main crystal axes, the initial (equilibrium) magnetization $M$ has all three non-zero components ($M_x, M_y, M_z$). Our analysis shows that the asymmetry of the trajectories is inextricably linked to the cubic symmetry of the photo-magnetic medium. In particular, we consider a cubic garnet crystal with an *m3m* point group symmetry. Assuming normal pump incidence so that $E_x \parallel [100], E_y \parallel [010], E_z = 0$, only three non-zero $\hat{\beta}$ components remain relevant in our case (see SM). The photo-magnetic excitation of spin dynamics is mediated by the effective photo-magnetic field $\mathbf{H_L} = -\frac{\partial \mathcal{H}}{\partial M}$. It then exerts a torque $T$ on magnetization, thus setting it into motion according to the LLG formalism: $\frac{\partial \mathbf{M}}{\partial t} = -\mathbf{M} \times \mathbf{H_L}$. Directly after the optical excitation, the out-of-plane laser-induced magnetization dynamics takes the following form:

$$\frac{\partial M_z}{\partial t} = \beta'(E_x^2 - E_y^2)M_x M_y + 2\beta'' E_x E_y (M_x^2 - M_y^2), \tag{2}$$

Here $\beta'$ and $\beta''$ are the linear combinations of the non-zero $\beta$ components: $\beta' \equiv (\beta_3 - \beta_1)$, $\beta'' \equiv 2\beta_2$ (see SM). Two main conclusions can be drawn from this result. First, consider the two initial magnetic states $M_8$ and $M_4$, as indicated in Fig. 1, that is, with identical in-plane magnetization components but opposite out-of-plane ones, $M_z^8 = -M_z^4$. Without the loss of generality, we refer to them as the "up" and "down" states, although the choice of up and down directions is arbitrary. Then, since it is even with respect to $M_z$, the normal component of the magnetization dynamics $\frac{\partial M_z}{\partial t}|_0$ will be the same for these two states. In other words, the "up" state will get the momentum towards the final "down" state of the switching process, whereas the "down" state will start its motion in the direction away from its destination. It is thus seen that the asymmetry of the switching trajectories is introduced immediately after the photo-magnetic excitation, and is governed by the tensor nature of the photo-magnetic effect and the cubic crystal symmetry.

Secondly, consider an excitation with linearly polarized light. Because the uniaxial anisotropy in our garnet film is much weaker than the cubic one, the equilibrium magnetization directions are close to the diagonals of the cubic unit cell. We thus can assume $M_x \approx M_y$ so that the first term in the normal torque component dominates. As such, the in-plane symmetry of the excitation is given by the $E_x^2 - E_y^2$ term, reducing to $E^2 \cos 2\varphi$, where $\varphi$ indicates the light polarization. Effectively, the in-plane symmetry of the photo-magnetic switching is lowered from *four-fold* to *two-fold*, which can be illustrated by rotating the polarization of the incident light from the [100] to the [010] direction. This leads to the reversal of the switching asymmetry: the normal component of the torque changes its sign, and thus the switching trajectories of the two considered initial magnetic states will be flipped. All these implications are in a striking agreement with the experimental findings.

**Modeling the trajectories of photo-magnetic switching.** Employing the extended LLG model of photo-magnetic switching[27] in which laser pulses act as a perturbation to the free energy density, we simulated the trajectories for the multi-states magnetization switching in a real YIG:Co(001) thin film. In the simulations, a tetragonal distortion of the cubic magnetic symmetry has been introduced (see SM). In Fig. 3a, the trajectories of magnetization switching induced by a normally incident laser pulse with the electric field $E$ are shown on the map, highlighting the polarization selectivity. The variations of the $M_z$ observed experimentally are summarized in Fig. 3b. Notably, the difference in the movement rates of the two simultaneously precessing magnetization vectors is enabled by the lack of coupling between them. This is further confirmed in the simulations where the angle between the two magnetization vectors (bottom panel) exhibits strong variations during their switching. Because magneto-dipole interaction in adjacent domains is too weak and manifests on the nanosecond timescale at best, the two magnetization vectors move independently.

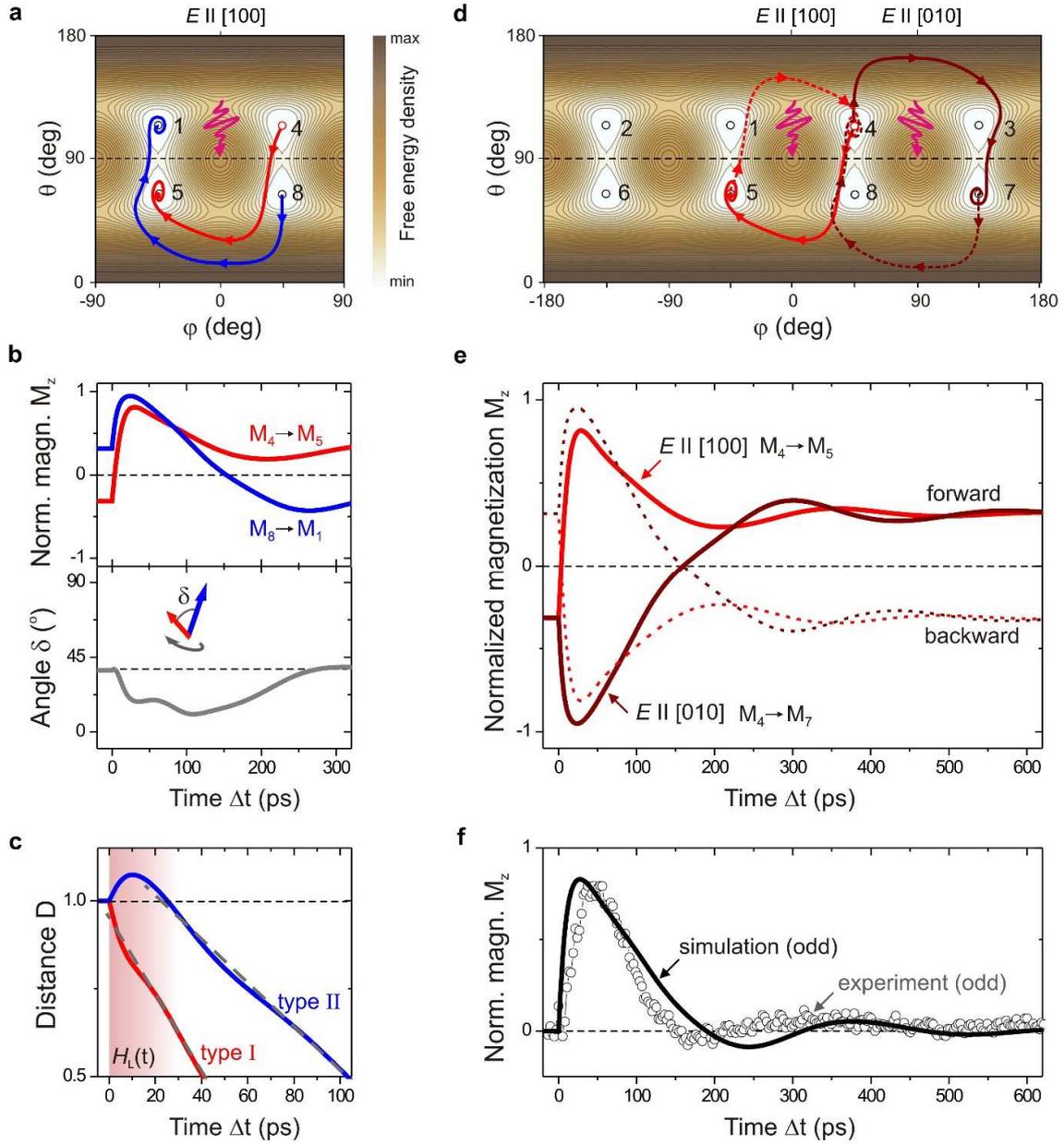

**Figure 3. Trajectories of magnetization dynamics on a fourfold-symmetric energy landscape. a** The simulated energy map with multi-states switching for different initial states **b** The trajectories of a normalized out-of-plane component $M_z$ with variations of the angle between moving vectors **c** The evolution of the distance to destination *D* highlighting type I and type II trajectories (dashed grey lines are the linear fitting). The red-shaded area shows the photo-induced anisotropy field $H_L(t)$ with a lifetime of about 20 ps. **d** The full simulated energy map showing different magnetization trajectories induced by the $E \parallel [100]$ or $E \parallel [010]$ light pulses. **e** Non-reciprocity of the all-optical switching caused by the change in the light pulse polarization plane (forward and backward switching). **f** Differential (odd) contribution to $E \parallel [100]$ and $E \parallel [010]$ trajectories responsible for the no-reciprocity compared with the experimental data.

It is apparent that the model captures the pronounced asymmetry of the switching trajectories, thus paving the way for further analysis. In particular, we turn to the apparent dissimilarity of the characteristic timescales when the magnetization takes either one of the two available switching routes. To quantify this, we calculated the evolution of the normalized distance to the route destination as:

$$D = \sqrt{\sum_i \left(M_i(t) - M_i^f\right)^2} / D_0. \tag{3}$$

where $M_i^f$ are the magnetization components of the final point on the trajectory and $D_0$ is the initial distance to the destination. A striking almost twofold difference in the movement rates along the two trajectories (Fig. 3c) corroborates their inequivalence and enables the qualitative distinction between the faster and slower (type I and type II as above) switching routes.

Lastly, the simulations reproduce the non-equivalence of the two types of trajectories in terms of transient variations of the normal magnetization component $\Delta M_z(\Delta t)$ (Fig. 3e). Depending on the polarization (parallel to either [100] or [010]) of the optical pump pulse, magnetization selects the fast or slow trajectory towards its destination, in agreement with the experimentally found non-reciprocity (cf. Fig. 2a). Similarly, the simulated odd component in Fig. 3f exhibits an oscillating character, again correlating well with the experimental observations and reinforcing our understanding of its origin in the sign of the initially produced photo-magnetic torque.

**Discussion**

The apparent non-reciprocity of the photo-magnetic switching routes is enabled by two main factors. Firstly, it is only possible when the switching is precessional in nature, as opposed to the thermal mechanisms accompanied by the magnetization quenching. Secondly, it requires an intricate interplay of the excitation tensor symmetry and magnetic energy landscape. In particular, in a uniaxial magnetic system, the observed asymmetry is impossible, which makes it difficult to imagine the discussed non-reciprocity in amorphous alloys where the dominant uniaxial contribution of magnetic anisotropy is often governed by the growth direction. Cubic magnetic crystals thus represent an attractive class of materials for exploring the richness of the non-thermal excitation of non-reciprocal magnetization dynamics. They constitute a promising playground for engineering magnetic energy landscapes in order to achieve faster and more efficient unidirectional switching. In particular, the change in cubic and uniaxial growth-induced anisotropy enables encoding materials with a spatial distribution of states with different energy thresholds, which can dramatically reduce the switching energy.

The approach developed in our work can be extended towards more sophisticated systems featuring additional external stimuli and physically rich interaction mechanisms. In particular, thermal demagnetization can be accounted for on equal footing by introducing Landau-Lifshitz-Bloch formalism. Moreover, there is tremendous potential in utilizing strain (either *dc* or phonon-induced) or auxiliary magnetic and electric fields to control the effective field of the anisotropy. Taking advantage of the extremely large phase space with these fields as control parameters, the switching between multiple domain states holds promise for the advancement of multi-level magnetic memories. In particular, the inequality of the switching trajectories indicates the possibility of simultaneously steering a set of magnetizations from one combination of states to another with a single stimulus, realizing a multi-dimensional magnetization switching. By encoding information not only in binary but in a set of magnetization states (e.g. logical combinations of 00, 01, 10, 11, etc.), a greater number of distinct levels can be achieved, allowing for higher data storage density.

**References**

1. Kirilyuk, A., Kimel, A. V. & Rasing, T. Ultrafast optical manipulation of magnetic order. *Rev. Mod. Phys.* **82**, 2731–2784 (2010).
2. Vedmedenko, E. Y. *et al.* The 2020 magnetism roadmap. *J. Phys. D. Appl. Phys.* **53**, (2020).
3. Kimel, A. V. & Li, M. Writing magnetic memory with ultrashort light pulses. *Nature Reviews Materials* vol. 4 189–200 (2019).
4. Kubacka, T. *et al.* Large-amplitude spin dynamics driven by a THz pulse in resonance with an electromagnon. *Science (80-. ).* **343**, 1333–1336 (2014).
5. Brataas, A., Kent, A. D. & Ohno, H. Current-induced torques in magnetic materials. *Nat. Mater.* **11**, 372–381 (2012).
6. Manipatruni, S. *et al.* Scalable energy-efficient magnetoelectric spin–orbit logic. *Nature* **565**, 35–42

(2019).

7. Scherbakov, A. V. *et al.* Coherent magnetization precession in ferromagnetic (Ga,Mn)As induced by picosecond acoustic pulses. *Phys. Rev. Lett.* **105**, 1–4 (2010).
8. Kovalenko, O., Pezeril, T. & Temnov, V. V. New concept for magnetization switching by ultrafast acoustic pulses. *Phys. Rev. Lett.* **110**, 1–5 (2013).
9. Vlasov, V. S. *et al.* Magnetization switching in bistable nanomagnets by picosecond pulses of surface acoustic waves. *Phys. Rev. B* **101**, 24425 (2020).
10. Afanasiev, D. *et al.* Laser excitation of lattice-driven anharmonic magnetization dynamics in dielectric FeBO3. *Phys. Rev. Lett.* **112**, 1–5 (2014).
11. Maehrlein, S. F. *et al.* Dissecting spin-phonon equilibration in ferrimagnetic insulators by ultrafast lattice excitation. *Sci. Adv.* **4**, (2018).
12. Stupakiewicz, A. *et al.* Ultrafast phononic switching of magnetization. *Nat. Phys.* **17**, 489–492 (2021).
13. Disa, A. S. *et al.* Polarizing an antiferromagnet by optical engineering of the crystal field. *Nat. Phys.* **16**, 937–941 (2020).
14. Radu, I. *et al.* Transient ferromagnetic-like state mediating ultrafast reversal of antiferromagnetically coupled spins. *Nature* **472**, 205–209 (2011).
15. Ostler, T. A. *et al.* Ultrafast heating as a sufficient stimulus for magnetization reversal in a ferrimagnet. *Nat. Commun.* **3**, (2012).
16. Chizhik, A., Davidenko, I., Maziewski, A. & Stupakiewicz, A. High-temperature photomagnetism in Co-doped yttrium iron garnet films. *Phys. Rev. B - Condens. Matter Mater. Phys.* **57**, 14366–14369 (1998).
17. Hansteen, F., Kimel, A., Kirilyuk, A. & Rasing, T. Femtosecond photomagnetic switching of spins in ferrimagnetic garnet films. *Phys. Rev. Lett.* **95**, 1–4 (2005).
18. Kalashnikova, A. M. *et al.* Impulsive generation of coherent magnons by linearly polarized light in the easy-plane antiferromagnet FeBO3. *Phys. Rev. Lett.* **99**, 1–4 (2007).
19. Atoneche, F. *et al.* Large ultrafast photoinduced magnetic anisotropy in a cobalt-substituted yttrium iron garnet. *Phys. Rev. B - Condens. Matter Mater. Phys.* **81**, 3–8 (2010).
20. Yoshimine, I. *et al.* Phase-controllable spin wave generation in iron garnet by linearly polarized light pulses. *J. Appl. Phys.* **116**, (2014).
21. Shelukhin, L. A. *et al.* Ultrafast laser-induced changes of the magnetic anisotropy in a low-symmetry iron garnet film. *Phys. Rev. B* **97**, (2018).
22. Stöhr, J. & Siegmann, H. C. *Magnetism. Journal of Physics A: Mathematical and Theoretical* (Springer Berlin Heidelberg, 2006).
23. Frej, A., Razdolski, I., Maziewski, A. & Stupakiewicz, A. Nonlinear subswitching regime of magnetization dynamics in photomagnetic garnets. *Phys. Rev. B* **107**, 134405 (2023).
24. Stupakiewicz, A., Szerenos, K., Afanasiev, D., Kirilyuk, A. & Kimel, A. V. Ultrafast nonthermal photo-magnetic recording in a transparent medium. *Nature* **542**, 71–74 (2017).
25. Görnert, P. *et al.* Co Containing Garnet Films with Low Magnetization. *Phys. Status Solidi* **74**, 107–112 (1982).
26. Maziewski, A. Unexpected magnetization processes in YIG + Co films. *J. Magn. Magn. Mater.* **88**, 325–342 (1990).
27. Stupakiewicz, A. *et al.* Selection rules for all-optical magnetic recording in iron garnet. *Nat. Commun.* **10**, (2019).
28. Zalewski, T. & Stupakiewicz, A. Single-shot imaging of ultrafast all-optical magnetization dynamics with a spatiotemporal resolution. *Rev. Sci. Instrum.* **92**, (2021).
29. Maziewski, A., Půst, L. & Görnert, P. Magnetometrical study of cobalt doped YIG garnet films. *J. Magn. Magn. Mater.* **83**, 87–88 (1990).

**Acknowledgments**. We acknowledge support from the grant of the Foundation for Polish Science POIR.04.04.00-00-413C/17-00 and the European Union's Horizon 2020 Research and Innovation Programme under the Marie Skłodowska-Curie grant agreement No 861300 (COMRAD). We thank Prof. A. Zvezdin for the fruitful discussions.

**Supplemental Material: Methods**

**Materials.** A monocrystalline cobalt-doped yttrium iron garnet (YIG:Co) thin film with composition $Y_2CaFe_{3.9}Co_{0.1}GeO_{12}$ was deposited on a (001) gadolinium gallium garnet substrate with a 4° miscut toward the [100] axis. YIG:Co exhibits optical transparency in the near-infrared range and has a static Faraday rotation of approximately 0.4° at the 650 nm wavelength. The $d$ = 7.5 μm-thick film was produced using a liquid phase epitaxy method [25]. The saturation magnetization at room temperature was 7 Gs and Gilbert damping $\alpha \approx 0.2$. The dominant cubic magnetocrystalline anisotropy of YIG:Co exhibits a negative first constant ($K_1 \approx -10^4$ erg cm$^{-3}$), while the growth-induced uniaxial anisotropy is smaller ($K_u \approx -3 \times 10^3$ erg cm$^{-3}$). This results in four easy magnetization axes of <111>-type directions in the garnet. In the absence of an external magnetic field, the garnet's domain structure exhibits both perpendicular and in-plane magnetization components, resulting in four distinct domain states[29]. The introduced miscut creates an imbalance in the distribution of domain states with the same in-plane magnetization component, making them easily distinguishable in a polarization microscope. Due to the distortion of cubic symmetry, two magnetization states are always spatially larger than the other two. These states are referred to as the large domain states $M_1$ and $M_8$, which correspond to the magnetizations along with directions close to [11-1] and [1-11], respectively (see Fig. 1a). The small labyrinth-like domain states $M_4$ and $M_5$ correspond to [1-1-1] and [111], respectively. All possible domain states can be induced by applying an in-plane magnetic field of approximately 2 mT directed along <110>-type axes[26].

**Experimental technique for time-resolved imaging of switching.** We employed the time-resolved (pump-probe) magneto-optical microscopy in the Faraday geometry. The initial ultrashort pulse train was generated using a Ti:Sapphire oscillator combined with an amplifier, resulting in a pulse duration of 35 fs and a base repetition rate of 1 kHz, enabling operation in the single-shot regime. The pump and probe beams passed through separate optical parametric amplifiers and were set to wavelengths of 1300 nm, for achieving the highest switching efficiency, and 650 nm, for optimizing Faraday rotation and absorption in the material, respectively. The probe beam, linearly polarized and defocused to illuminate a sufficiently large sample area, was impinged on the sample at normal incidence. The pump beam was focused on the sample into a spot of approximately 120 μm in diameter with a fluence of $F$ = 50 mJ cm$^{-2}$. The polarization plane of the pump beam was aligned with the [100] or [010] direction of the garnet crystallographic axis and controlled by a half-wave plate. All measurements were performed at room temperature in the absence of an external magnetic field.

We obtained a stack of time-resolved magneto-optical images[28], showing the spatial and temporal magnetization changes in the four types of magnetic domains in YIG:Co. The measurement sequence for each image in the stack involved capturing a background image (illuminated solely by the probe beam), capturing a dynamic image (both pump and probe beams illuminating the sample), and resetting the domain structure to its initial state. The time delay $\Delta t$ was controlled by varying the optical path of the pump beam utilizing a motorized translation stage. To enhance the signal-to-noise ratio and isolate the pump-induced changes, we subtracted the background image from each dynamic pump-probe image, creating a stack of differential images. This relative change provides the time-dependent evolution of the magnetization projection $\Delta M_z$ along the [001] axis which can be normalized to the static Faraday rotation. Through analysis of the temporal behavior of spatially defined regions of interest (RoI) in the image corresponding to one of the domain states, we were able to distinguish the magnetization trajectories in each domain independently. Thus, selecting the RoI position and size allows probing spatially averaged $\Delta M_z$ changes, enabling precise separation of the magnetization dynamics in different domain states. To account for the spatial dependence of the photo-induced anisotropy strength originating in the Gaussian pump beam profile (Fig. 4a), the analyzed RoIs in the domains were manually chosen to have the same size and lie close to each other.

Taking into account the studied garnet absorption of 12% ($a$ = 0.12) at 1300 nm, a molar mass of $m$ = 706 g mole$^{-1}$, a heat capacity $C$ = 430 J mole$^{-1}$, and mass density $\rho$ = 7.12 g cm$^{-3}$ we estimated the local laser-induced heating of the sample as $q = aF = \Delta T C \rho d / m$. The averaged measured pump fluence was 50 mJ cm$^{-2}$, which corresponds to a peak value of 100 mJ cm$^{-2}$ for a centrally located RoI. This results in an approximate temperature change of 4 K (see Fig. 4a).

In addition, we compared the dynamics of magnetization precession using our method of spatially-resolved time-resolved imaging (see Fig. 4d) to that obtained in the conventional pump-probe technique (with spatial averaging) (see Fig. 2b). Both measurement methods yield highly consistent results.[28]

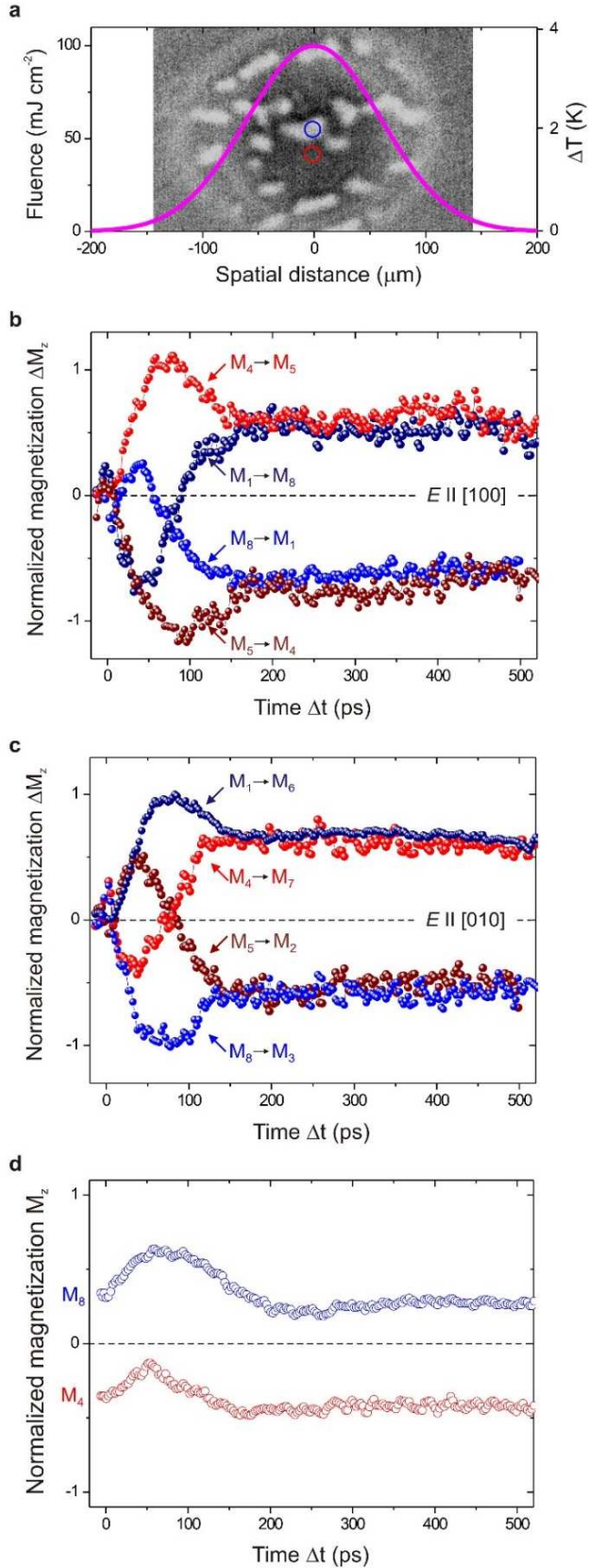

**Figure 4. Dynamics of magnetization reversal between all states in YIG:Co. a** The pump intensity profile measured with a beam profiler resulting in the spatially inhomogeneous field of photo-magnetic anisotropy. The red and blue circles indicate RoIs in the two domains chosen for the subsequent analysis. The right vertical axis indicates calculated spatially-resolved laser-induced heating. **b** The normal component of magnetization $\Delta M_z$ for all different magnetic states illuminated with E || [100] or **c** E || [010] pump polarization. **d** The magnetization precession in the initial states ($M_4$ and $M_8$) at a pump fluence of 10 mJ cm$^{-2}$ below the switching threshold.

**Photo-magnetic symmetry analysis.** We consider a reference frame aligned with the main crystal axes. The analysis is substantially simplified by the normal incidence of the pump beam, that is, parallel to the (001) direction. In the initial state, magnetization **M** has all three non-zero components $(M_x, M_y, M_z)$. Owing to the cubic $m3m$ symmetry of the non-distorted garnet and setting $E_x \parallel$ (100), $E_y \parallel$ (010), $E_z = 0$, we can outline the following non-zero components of the photo-magnetic susceptibility tensor $\beta$:

$$\beta_1 \equiv \beta_{xxxx} = \beta_{yyyy};$$

$$\beta_2 \equiv \beta_{xyxy} = \beta_{yxyx} = \beta_{yxxy} = \beta_{xyyx};$$

$$\beta_3 \equiv \beta_{xxyy} = \beta_{yyxx} = \beta_{xxzz} = \beta_{yyzz}.$$

At the initial stage directly after the laser excitation, under the action of the effective photo-magnetic field $H_{L,l} = -\beta_{ijkl} E_i E_j M_k$ the magnetization dynamics takes the following shape:

$$\frac{\partial M_x}{\partial t}\bigg|_0 = (\beta_3 - \beta_1) E_y^2 M_y M_z - 2\beta_2 E_x E_y M_x M_z$$

$$\frac{\partial M_y}{\partial t}\bigg|_0 = -(\beta_3 - \beta_1) E_x^2 M_x M_z + 2\beta_2 E_x E_y M_y M_z$$

$$\frac{\partial M_z}{\partial t}\bigg|_0 = (\beta_3 - \beta_1)(E_x^2 - E_y^2) M_x M_y + 2\beta_2 E_x E_y (M_x^2 - M_y^2)$$

Our analysis reveals that the non-reciprocal dynamics of magnetization is triggered by a sign change of the photomagnetic torque acting on the cubic magnetic symmetry in the garnet.

**Numerical simulation of magnetization trajectories.** For the numerical simulations, we consider the free energy consisting of the cubic and effective (including demagnetization term) uniaxial anisotropy terms $W_c$ and $W_{\text{eff}} = W_u - 2\pi M_s^2 \sin^2\theta$ respectively, and of the photoinduced magnetic anisotropy[23] with the field $H_L$.

$$W_{tot} = \frac{K_1}{4}[(\sin 2\theta)^2 + (\sin\theta)^4 (\sin 2\varphi)^2] + K_{\text{eff}}(\sin\theta)^2 + \beta_{ijkl} E_i E_j M_k M_l$$

The simulations of magnetization trajectories during the photo-magnetic reversal in a YIG:Co film with a $4mm$ symmetry have been performed by numerically solving the LLG equation in a spherical $(\theta, \varphi)$ coordinates:

$$\dot{\theta} = -\frac{\gamma K_1}{2M_s}(\sin\theta)^3 \sin 4\varphi + bFe^{-t/\tau}\sin\theta \sin 2\varphi - \alpha\sin\theta\,\dot{\varphi},$$

$$\sin\theta\,\dot{\varphi} = \frac{\gamma K_{\text{eff}}}{M_s}\sin 2\theta + \frac{\gamma K_1}{2M_s}[\sin 2\theta\,(\sin\theta)^2(\sin 2\varphi)^2 + \sin 4\theta]$$

$$+ \frac{F}{2}e^{-t/\tau}(1 + b\cos 2\varphi)\sin 2\theta + \alpha\dot{\theta},$$

where the laser fluence is defined by $F = \gamma|E_0|^2/M_s$, and $\gamma$ is the gyromagnetic ratio. The lifetime of the photo-induced anisotropy was $\tau = 20$ ps[24]. In Figure 5, we demonstrate the multi-state magnetization switching at $F = 60$, in a good agreement with experimental results (see Fig. 1d and Fig. 2a). At low excitation levels ($F = 10$, below the switching threshold as shown in Fig. 5c) small-angle precession is observed, in agreement with the experimental results (see Fig. 4d).

In the modeling, we adjusted the parameter $b$ to obtain a better agreement with the experimental results of multi-state switching. We note that the tetragonal distortion of the cubic garnet lattice is characterized by ($b$-1), and in a perfect cubic symmetry $b = 1$. Previously,[27] for the single-state switching we considered the parameter *b* in the range 0.3<|*b*|<0.5. Here, we set |*b*|= 1.6 and obtained a good correspondence with the results of the multi-state experiment.

The azimuthal angle $\varphi$ almost does not change in the beginning, in contrast to the polar angle $\theta$. Strong changes of $\theta$ lead to the fact that the switching trajectory approaches the pole ($|M_z|\sim 1$), (see Fig. 5), in agreement with the experimental data in Fig. 4.

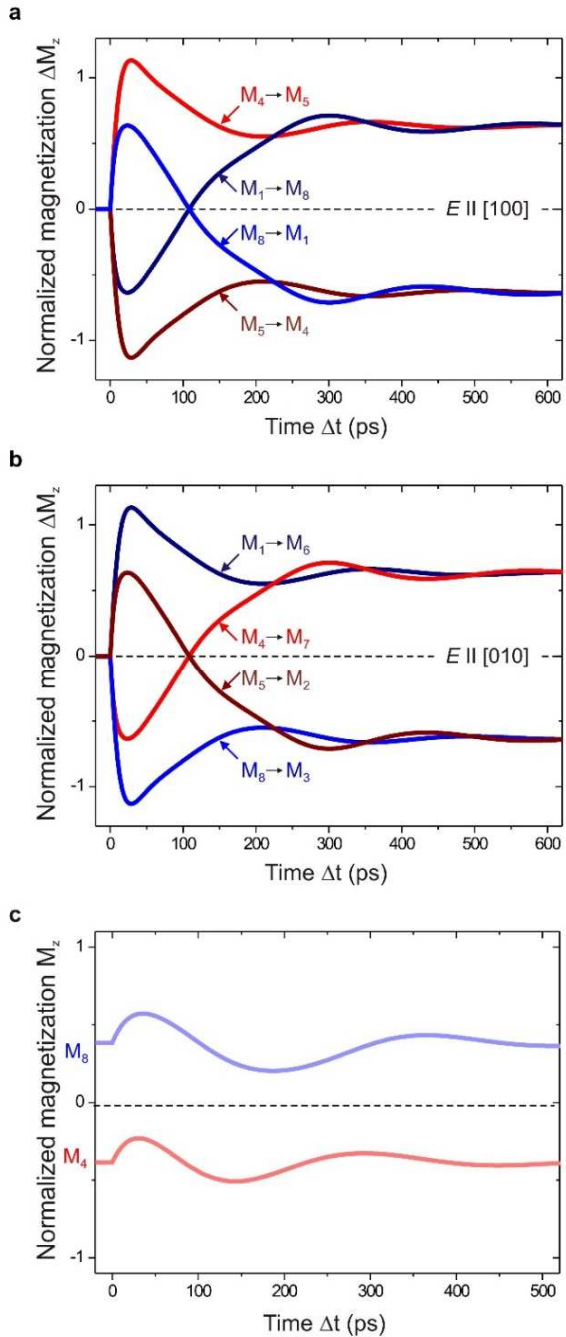

**Figure 5. Simulation for all states** (as in Figure 4) for **a** E || [100] and **b** E || [010] pump polarization with an intensity $F$ = 60. **c** The simulation of magnetization precession for an intensity